\documentclass{article}

\usepackage{PRIMEarxiv}

\usepackage[utf8]{inputenc} 
\usepackage[T1]{fontenc}    
\usepackage{hyperref}       
\usepackage{url}            
\usepackage{booktabs}       
\usepackage{amsfonts}       
\usepackage{nicefrac}       
\usepackage{microtype}      
\usepackage{lipsum}
\usepackage{fancyhdr}       
\usepackage{graphicx}       
\graphicspath{{media/}}     

\usepackage[numbers]{natbib}
\usepackage{multirow}
\usepackage{graphicx}
\usepackage{paralist}

\usepackage{amsmath} 
\usepackage{cleveref}
\crefname{figure}{Fig.}{Figs.}
\crefname{section}{Sect.}{Sects.}

\usepackage[linesnumbered,algoruled,boxed,lined]{algorithm2e}
\SetKwRepeat{Do}{do}{while}
\usepackage{balance}
\usepackage{xspace}
\usepackage{graphicx}
\usepackage{subcaption}
\AtBeginDocument{%
}

\newcommand{\fitnessDistWPs}{\ensuremath{\mathit{fit_{distWPs}}}\xspace}
\newcommand{\fitnessUnstable}{\ensuremath{\mathit{fit_{unstable}}}\xspace}
\newcommand{\vessel}{\ensuremath{\mathcal{V}}\xspace}
\newcommand{\simulate}{\ensuremath{\mathit{simulate}}\xspace}
\newcommand{\pathVessel}{\ensuremath{\mathit{p}}\xspace}
\newcommand{\subpathVessel}{\ensuremath{\mathit{sp}}\xspace}
\newcommand{\pathX}[1]{\ensuremath{x^{p}_{#1}}\xspace}
\newcommand{\pathY}[1]{\ensuremath{y^{p}_{#1}}\xspace}
\newcommand{\lengthPath}{\ensuremath{\mathit{K}}\xspace}
\newcommand{\deltaSearchSpace}{\ensuremath{\Delta}\xspace}
\newcommand{\numWaypoints}{\ensuremath{\mathit{N}}\xspace}

\newcommand{\numOfMutParams}{\ensuremath{\mathit{numOfMutParams}}\xspace}

\newcommand{\waypointGeneric}{\ensuremath{\mathit{wp}}\xspace}
\newcommand{\setWaypointsGeneric}{\ensuremath{\mathcal{WP}}\xspace}
\newcommand{\waypointGenericIndex}[1]{\ensuremath{\waypointGeneric_{#1}}\xspace}

\newcommand{\waypointOrig}{\ensuremath{\waypointGeneric^o}\xspace}
\newcommand{\setWaypointsOrig}{\ensuremath{\setWaypointsGeneric_o}\xspace}
\newcommand{\waypointOrigIndex}[1]{\ensuremath{\waypointOrig_{#1}}\xspace}
\newcommand{\waypointOrigX}[1]{\ensuremath{v^{o}_{x, #1}}\xspace}
\newcommand{\waypointOrigY}[1]{\ensuremath{v^{o}_{y, #1}}\xspace}

\newcommand{\waypointPrime}{\ensuremath{\waypointGeneric^{\prime}}\xspace}
\newcommand{\setWaypointsPrime}{\ensuremath{\setWaypointsGeneric^{\prime}}\xspace}
\newcommand{\waypointPrimeIndex}[1]{\ensuremath{\waypointPrime_{#1}}\xspace}

\newcommand{\waypointMod}{\ensuremath{\waypointGeneric}\xspace}
\newcommand{\setWaypointsMod}{\ensuremath{\setWaypointsGeneric}\xspace}
\newcommand{\waypointModIndex}[1]{\ensuremath{\waypointMod_{#1}}\xspace}

\newcommand{\rSwitch}{\ensuremath{\mathit{minWPdist}}\xspace}
\newcommand{\Atwelve}{\ensuremath{\hat{A}}\textsubscript{12}\xspace}
\newcommand{\approach}{\texttt{WPgen}\xspace}
\newcommand{\approachSeed}{\texttt{WPgen}$_{\text{\texttt{seed}}}$\xspace}
\newcommand{\randomSearch}{\texttt{RS}\xspace}
\newcommand{\approachComb}{\texttt{WPgen}$_{\text{\texttt{comb}}}$\xspace}
\newcommand{\approachRnd}{\texttt{WPgen}$_{\text{\texttt{rnd}}}$\xspace}
\newcommand{\searchVariables}{\ensuremath{\overline{k}}\xspace}
\newcommand{\searchVariable}{\ensuremath{k}\xspace}
\newcommand{\searchValue}{\ensuremath{v}\xspace}
\newcommand{\searchValues}{\ensuremath{\overline{v}}\xspace}
\newcommand{\modPoint}{\ensuremath{\mathit{mp}}\xspace}
\newcommand{\randZeroOne}{\ensuremath{\mathit{rand}}\xspace}
\newcommand{\randomNumber}{\ensuremath{\mathit{randomNumber}}\xspace}
\newcommand{\resmus}{\texttt{Remus100}\xspace}
\newcommand{\nspauv}{\texttt{Nspauv}\xspace}
\newcommand{\mariner}{\texttt{Mariner}\xspace}
\newcommand{\EUProject}{{\tt RoboSAPIENS}\xspace}
\newcommand{\neglibleStrength}{\texttt{negligible}\xspace}
\newcommand{\smallStrength}{\texttt{small}\xspace}
\newcommand{\mediumStrength}{\texttt{medium}\xspace}
\newcommand{\largeStrength}{\texttt{large}\xspace}
\newcommand{\noDifference}{\texttt{ND}\xspace}
\newcommand{\signLevel}{\ensuremath{\alpha}\xspace}

\usepackage[most]{tcolorbox}
\newcommand{\conclusion}[1]{\begin{tcolorbox}[size=title, colframe=white, width=1\linewidth, colback=gray!20, breakable=true]
#1
\end{tcolorbox}}

\title{Search-based Generation of Waypoints for Triggering Self-Adaptations in Maritime Autonomous Vessels}
  
\author{
  Karoline Nylænder \\
  Simula Research Laboratory and\\
  Oslo Metropolitan University \\
  Oslo, Norway \\
  \texttt{karolinen@simula.no} \\
  \And
  Aitor Arrieta \\
  Mondragon University \\
  Mondragon, Spain \\
  \texttt{aarrieta@mondragon.edu} \\
  \and
  \textbf{Shaukat Ali} \\
  Simula Research Laboratory\\
  Oslo, Norway \\
  \texttt{shaukat@simula.no} \\
  \And
  Paolo Arcaini\\
  National Institute of Informatics \\
  Tokyo, Japan \\
  \texttt{aarrieta@mondragon.edu} \\
}

\begin{document}
\maketitle

\begin{abstract}
Self-adaptation in maritime autonomous vessels (AVs) enables them to adapt their behaviors to address unexpected situations while maintaining dependability requirements. During the design of such AVs, it is crucial to understand and identify the settings that should trigger adaptations, enabling validation of their implementation. To this end, we focus on the navigation software of AVs, which must adapt their behavior during operation through adaptations. AVs often rely on predefined waypoints to guide them along designated routes, ensuring safe navigation. We propose a multi-objective search-based approach, called WPgen, to generate minor modifications to the predefined set of waypoints, keeping them as close as possible to the original waypoints, while causing the AV to navigate inappropriately when navigating with the generated waypoints. WPgen uses NSGA-II as the multi-objective search algorithm with three seeding strategies for its initial population, resulting in three variations of WPgen. We evaluated these variations on three AVs (one overwater tanker and two underwater). We compared the three variations of WPgen with Random Search as the baseline and with each other. Experimental results showed that the effectiveness of these variations varied depending on the AV. Based on the results, we present the research and practical implications of WPgen. 
\end{abstract}


\section{Introduction}

Self-adaptation enables systems to handle unforeseen situations while maintaining dependable operation. Implementing such self-adaptation in robotic software is a key objective of a large European project (\EUProject), within the framework of which the work reported in this paper is being conducted. The project focuses on extending the well-established MAPE-K self-adaptation loop~\cite{MAPE-K} and applying it to various real-world industrial robotic software systems, enabling them to adapt effectively to unexpected situations. This paper focuses on one real-world application provided by the project, i.e., maritime autonomous vessels (AVs).

This paper considers the {\it navigation software} of the AV, as this is one of the most critical of its software components. This software typically relies on a navigation method using waypoints to help the AV accurately and safely generate its path to the intended destination~\cite{Handbook}. When an AV follows its generated path, various conditions, such as environmental factors, sensor errors, or software malfunctions, can require it to adapt its behavior. Another type of event that may require adaptation is the presence of waypoints different from those the AV was supposed to follow. Indeed, the navigation software may not be properly tuned to handle these waypoints, and it may require an adaptation. This paper explicitly focuses on identifying waypoints that should trigger adaptations to prevent navigation software from malfunctioning. In particular, we propose a multi-objective search approach, called \approach, to generate a set of waypoints by introducing minor modifications to an existing set of waypoints that should trigger adaptations. To guide the search, we defined two objectives, i.e., minimizing the variations of the original waypoints and maximizing the path length to generate waypoints. We employed NSGA-II as the multi-objective search algorithm in \approach and implemented three strategies to seed its initial population, resulting in three variations of \approach.

Our work focuses on generating minor variations to predefined waypoints and is thus related to test scenario generation for AVs. To this end, scenario-based simulation testing is commonly used~\cite{VVANavigation,ScenariNNTest,DLforNavigation,SMTwGaussian,RTforAVTest,SearchApproachMuliShipCollision}, with a specific focus on collision detection. Such methods often generate a limited number of scenarios (often manually created) under fixed settings~\cite{VVANavigation}. Recent approaches have begun exploring techniques, such as machine learning~\cite{ScenariNNTest,DLforNavigation}, random testing~\cite{RTforAVTest}, and search approaches~\cite{SearchApproachMuliShipCollision} for test automation. In contrast, \approach focuses on generating minor variations in given waypoints (i.e., tests) that should trigger adaptations in AVs. Moreover, \approach is a real-world application of search-based software engineering (SBSE)~\cite{SSBSE} in the maritime domain.

To evaluate \approach, we used the Marine Systems Simulator (MSS) \cite{MSSsimulator}, a widely used maritime simulator for the design, development, and testing of marine applications~\cite{czaplewski2016vessel}. We chose three AV models (one overwater tanker, and two underwater) and their navigation software that uses waypoint-based navigation. We compared the three variations of \approach with Random Search (\randomSearch) as baseline. Experimental results showed that the effectiveness of \approach depends on the AV. Moreover, we study characteristics of various unexpected paths generated by \approach. Such characterization helps determine which adaptations to trigger for which path types. Finally, we present research and practical implications of \approach.

\section{Real-world context}\label{sec:realWorldContext}

\EUProject~\cite{RoboSAPIENS} is a large European initiative focused on developing self-adaptive software for robotic applications across several domains, including maritime, manufacturing, and service. The project aims to develop novel methods for testing and implementing robotic software adaptations in unforeseen situations, enabling robots to handle such situations during real-world operations. The project collaborates with research organizations and industrial partners, providing real-world robotic software use cases for evaluating the developed methods. This paper focuses on our use case from the maritime domain, i.e., AVs. These AVs operate in the marine environment and are subject to rapid changes, e.g., in wind and ocean currents, resulting in uncertain AV behavior that can lead to unsafe situations. 

\begin{figure}[!tb]
\centering
\includegraphics[width=0.6\columnwidth]{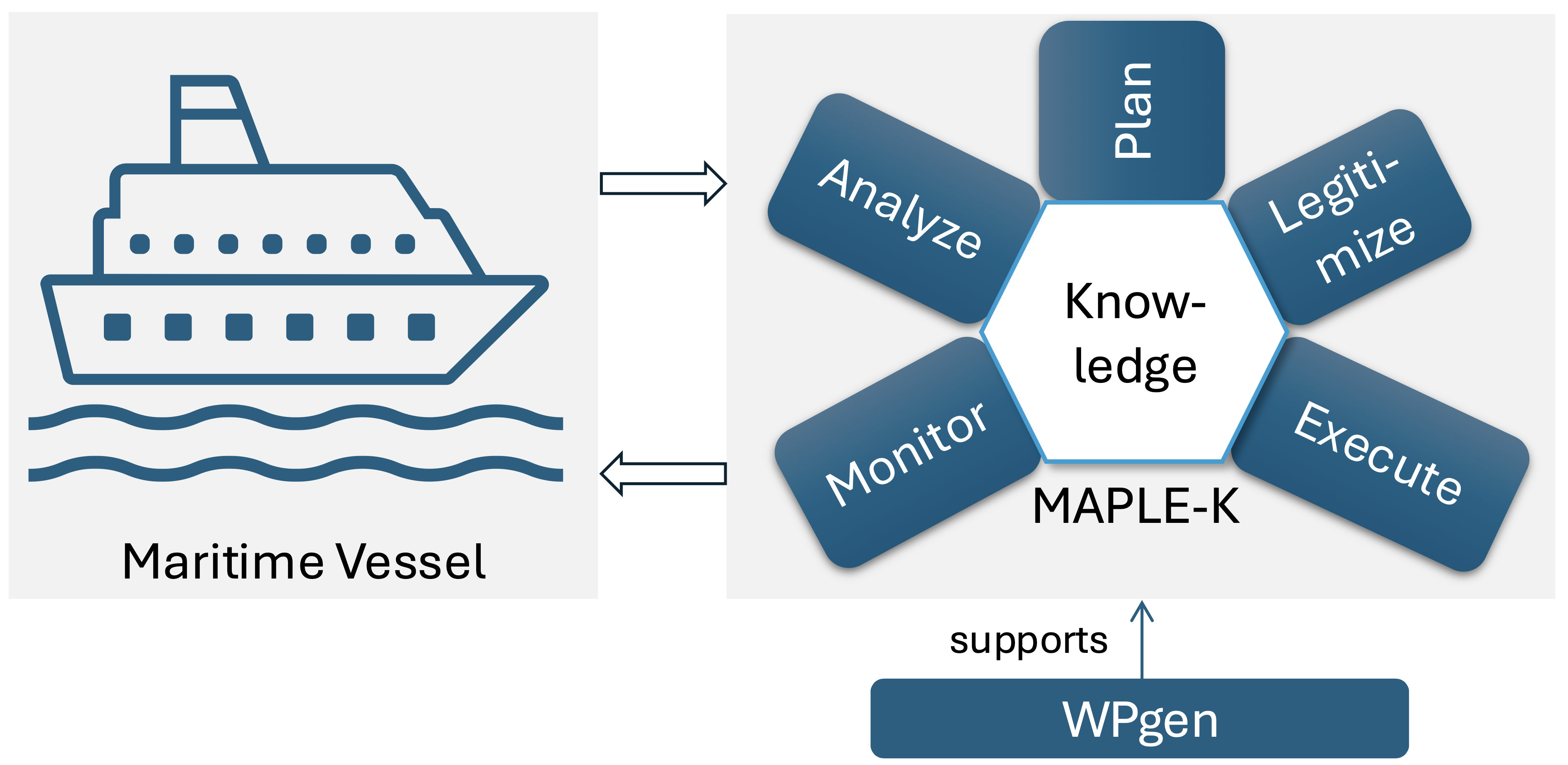}
\caption{\approach in the Context of the MAPLE-K loop}
\label{fig:context}
\end{figure}

The long-term goal of the project is to implement self-adaptation using the MAPLE-K~\cite{MAPLE-K} (see \Cref{fig:context}) loop—an extension of the well-established MAPE-K~\cite{MAPE-K} loop—to enable AVs to adapt their behavior in the presence of uncertainties and prevent potentially unsafe situations.
A typical MAPE-K loop works as follows: the \textbf{M}onitor component continuously monitors the state of the system (e.g., an AV), which is then analyzed by the \textbf{A}nalyze component. If the analysis indicates that the system requires adaptation, the \textbf{P}lan component devises a plan, which is subsequently executed by the \textbf{E}xecute component to perform the adaptation on the system. MAPLE-K extends MAPE-K by introducing the \textbf{L}egitimate component, which verifies whether the adaptation meets all safety requirements. All components contribute to the \textbf{K}nowledge component, where relevant information is stored. At the time of development of the self-adaptive software, it is essential to find settings that should trigger adaptations. This is needed to ensure that not only the triggered adaptations are correct but also that the adaptations' implementations are correct, preventing the AVs from going into unsafe situations.

The project is currently in its early phase, focusing on implementing the MAPLE-K loop in its robotic software, while our work, in parallel, aims to develop an approach to test the MAPLE-K loop. Specifically, it generates modified waypoints that should trigger relevant adaptations, thereby assessing whether the Plan and Legitimate components can successfully select and verify appropriate adaptations, respectively.

\section{Approach}\label{sec:approach}

\subsection{Approach Overview} \label{sec:overview}

We represent an AV and its associated {\it navigation software} as \vessel. In particular, we consider {\it waypoint-based} navigation software that guides the AV through a set of predefined {\it waypoints}. A \textit{waypoint} \waypointGeneric is a coordinated destination point. For an overwater AV, a waypoint is defined in a North-East coordinate system with two dimensions (i.e., $\waypointGenericIndex{i} = (x_i,$ $y_i)$). Similarly, for an underwater AV, the waypoint is defined in a North-East-Down coordinate system with 3 dimensions (i.e., $\waypointGenericIndex{i} = (x_i,$ $y_i,$ $z_i)$). A path is defined as a set of \numWaypoints waypoints $\setWaypointsGeneric = \{\waypointGenericIndex{1},$ $\ldots,$ $\waypointGenericIndex{\numWaypoints}\}$ that the AV \vessel must navigate through.

Waypoints placement must ensure that the AV has sufficient space to navigate smoothly between them. Specifically, each pair of consecutive waypoints \waypointGenericIndex{i} and \waypointGenericIndex{i+1} cannot be closer than \rSwitch. The value of \rSwitch depends on the specific AV.

Our approach \approach generates waypoints representing possible alterations of the original set of waypoints $\setWaypointsOrig = \{\waypointOrigIndex{1},$ $\ldots,$ $\waypointOrigIndex{\numWaypoints}\}$ that the AV \vessel is originally supposed to follow. These variations could occur during navigation due to environmental disturbances such as wind and sea current, or obstacles that block the planned path.

The goal of \approach is to find a set of modified waypoints that are challenging for the current navigation software \vessel, i.e., that lead to obtaining {\it unstable} paths of the AV. These sets of waypoints will be used to implement a self-adaptive navigation software that can, at runtime, safely self-adapt to these runtime variations. Note that implementing such self-adaptive navigation software is not part of this work. This implementation is being carried out by our industrial partners in \EUProject, in parallel with this work.

\approach is cast as a multi-objective optimization problem that is described in detail in the following sections.

\subsection{Solution Encoding}\label{sec:solution}
In \approach, the {\it search variables} \searchVariables represent a set of $\numWaypoints - 1$ consecutive waypoints that define the modified set of waypoints the AV \vessel must follow during navigation. We do not search for the first waypoint that is the same as that of \setWaypointsOrig (i.e., \waypointOrigIndex{1}), as it is the vessel's current starting point and it must be common to all the \vessel's paths.

For each searched waypoint, there are two or three variables, depending on whether the search is conducted for an overwater or an underwater AV. For simplicity, in the following, we consider an overwater AV (the formalization for an underwater AV is similar); so, the search variables are $\searchVariables = \{\searchVariable_{2, x},$ $\searchVariable_{2, y},$ $\ldots,$ $\searchVariable_{\numWaypoints, x},$ $\searchVariable_{\numWaypoints, y}\}$, where $(\searchVariable_{i, x},$ $\searchVariable_{i, y})$ defines the $i$th waypoint.

A {\it solution} (or {\it individual}) is a concrete assignment $\searchValues = \{\searchValue_{x,2},$ $\searchValue_{y,2},$ $\ldots,$ $\searchValue_{x,\numWaypoints},$ $\searchValue_{y,\numWaypoints}\}$, that defines the modified waypoints $\waypointModIndex{2} = (\searchValue_{x,2}, \searchValue_{y,2})$, $\ldots$, $\waypointModIndex{\numWaypoints} = (\searchValue_{x,\numWaypoints}, \searchValue_{y,\numWaypoints})$. So, together with the starting waypoint \waypointOrigIndex{1}, these define the set of waypoints $\setWaypointsMod = \{\waypointOrigIndex{1},$ $\waypointModIndex{2},$ $\ldots,$ $\waypointModIndex{\numWaypoints}\}$ that will be executed in the AV's simulator MSS as test for the navigation software.

Since the generated set of waypoints \setWaypointsMod must be similar to the original set of waypoints $\setWaypointsOrig = \{\waypointOrigIndex{1},$ $\ldots,$ $\waypointOrigIndex{\numWaypoints}\}$, the search space of variables \searchVariables is defined in the neighborhood of the values of $\setWaypointsOrig =$ $\{(\waypointOrigX{1}, \waypointOrigY{1}),$ $\ldots,$ $(\waypointOrigX{\numWaypoints}, \waypointOrigY{\numWaypoints})\}$. Specifically, for each pair of variables $\searchVariable_{i,x}$, $\searchVariable_{i,y}$ (with $i \in \{2, \ldots, \numWaypoints\}$), their search spaces are $[\waypointOrigX{i} - \deltaSearchSpace, \waypointOrigX{i} + \deltaSearchSpace]$ and $[\waypointOrigY{i} - \deltaSearchSpace, \waypointOrigY{i} + \deltaSearchSpace]$, where \deltaSearchSpace is a hyperparameter of the approach.

As explained in \Cref{sec:overview}, consecutive waypoints must be at least at a distance of \rSwitch. Solutions generated by \approach could violate such constraint. Therefore, for each individual, \approach checks if the constraint is satisfied and, if it is not, it discards the individual (i.e., following a death penalty approach~\cite{surveyConstraints}).

\subsection{Initial population}
\label{sec:initial_pop}

At the beginning of the search, \approach must be initialized with an initial population. Prior works have shown that seeding the initial population is beneficial in terms of both effectiveness as well as efficiency of the algorithm when generating and selecting test cases~\cite{fraser2012seed,arrieta2023some,rojas2016seeding}. Because of this, \approach can be configured to use two different seeding strategies and a traditional fully random strategy, described in the following.

\subsubsection{\approachSeed{} -- Closeness seed}
This first strategy takes as input the original set of waypoints \setWaypointsOrig and applies minor mutations by following Algorithm~\ref{alg:solmut}.
\begin{algorithm}[!tb]
\caption{\approachSeed and \approachComb{} -- Generation of initial individual}\label{alg:solmut}
\small
\KwIn{$\setWaypointsOrig = \{\waypointOrigIndex{1}, \ldots, \waypointOrigIndex{\numWaypoints}\}$ (original waypoints)}
\KwOut{$\setWaypointsPrime = \{\waypointPrimeIndex{1}, \ldots, \waypointPrimeIndex{\numWaypoints}\}$ (modified waypoints)}
\numOfMutParams $\leftarrow$ 0;\\
\setWaypointsPrime $\leftarrow$ \setWaypointsOrig;\\
\Do{$p < 0.5^{\numOfMutParams}$}{
\modPoint $\leftarrow$ $\mathit{randomNumber}(\text{2 to }\numWaypoints)$; //select waypoint \waypointPrimeIndex{\modPoint}\label{line:selectWaypoint}\\
$\waypointPrimeIndex{\modPoint} \leftarrow \left(\begin{array}{l}
\randomNumber((\waypointOrigX{i} - \deltaSearchSpace)\text{ to }(\waypointOrigX{i} + \deltaSearchSpace))\\
\randomNumber((\waypointOrigY{i} - \deltaSearchSpace)\text{ to }(\waypointOrigY{i} + \deltaSearchSpace))
\end{array}\right)$;\label{line:mutateWaypoint}\\
\numOfMutParams $\leftarrow$ \numOfMutParams + 1; \\
$p$ $\leftarrow$ \randZeroOne(); //\textit{returns random value between 0 and 1}\\ 
}
\Return \setWaypointsPrime;
\end{algorithm}
Specifically, the strategy aims at mutating (in the allowed search space) at least one randomly selected waypoint of the initial solution, as shown in Lines~\ref{line:selectWaypoint}-\ref{line:mutateWaypoint}.
As the number of mutated variables (i.e., \numOfMutParams) increases, the chances for mutating a new variable decreases.

\subsubsection{\approachComb{} -- Combined} The second strategy takes as input the original waypoints \setWaypointsOrig used for testing the AV and applies minor mutations (following Algorithm~\ref{alg:solmut}) to half of the solutions in the population. The other half of the population is generated entirely randomly (still in the search space of search variables), promoting, in this way, diversity in the population.

\subsubsection{\approachRnd{} -- Random}
The last strategy follows the traditionally used fully random generation of the entire population. Since we want to produce waypoints close to the set of original waypoints \setWaypointsOrig, one of the individuals in the population is \setWaypointsOrig.

\subsection{Objective Functions}\label{sec:objectiveFunctions}

Given a set of waypoints \setWaypointsMod obtained from an individual \searchValues (see \Cref{sec:solution}), the approach simulates the navigation of AV \vessel over \setWaypointsMod, and obtains the path \pathVessel followed by \vessel, i.e.,
\begin{equation}
\pathVessel = \{(\pathX{1}, \pathY{1}), \ldots, (\pathX{\lengthPath}, \pathY{\lengthPath})\} = \simulate(\vessel, \setWaypointsMod)
\end{equation}

If the AV managed to follow all the waypoints in \setWaypointsMod, \pathVessel can be split in $\numWaypoints - 1$ {\it sub-paths} $\subpathVessel_1$, \ldots, $\subpathVessel_{\numWaypoints - 1}$, i.e., the sections of \pathVessel between each consecutive pair of waypoints. Note that the AV may fail to follow all the waypoints: in that case, the number of sub-paths is lower than $\numWaypoints - 1$.

\approach has two competing objectives that capture the overall goal of the approach, i.e., finding the minimal variations of the original waypoints \setWaypointsOrig that are challenging for the AV.

The first objective aims at minimizing the distance between $\setWaypointsOrig =$ $\{(\waypointOrigX{1}, \waypointOrigY{1}),$ $\ldots,$ $(\waypointOrigX{\numWaypoints}, \waypointOrigY{\numWaypoints})\}$ and the set of waypoints $\setWaypointsMod = \{(\searchValue_{x,1}, \searchValue_{y,1}),$ $\ldots,$ $(\searchValue_{x,\numWaypoints},$ $\searchValue_{y,\numWaypoints})\}$ obtained from an individual\footnote{With an abuse of notation, we compute the fitness on the set of waypoints \setWaypointsMod obtained from an individual \searchValues.}:
\begin{equation}\label{eq:fitnessDistWPs}
\fitnessDistWPs(\setWaypointsMod) = \sqrt{\sum_{i = 1}^{\numWaypoints}((\waypointOrigX{i} - \searchValue_{x,i})^2 + (\waypointOrigY{i} - \searchValue_{y,i})^2)}
\end{equation}

The second objective aims at finding {\it unstable paths}. Intuitively, a path is more unstable if it largely deviates from the optimal path composed of stable segments between consecutive waypoints. So, the fitness function that must be maximized is defined as follows:
\begin{equation}\label{eq:fitnessLengthPath}
\fitnessUnstable(\setWaypointsMod) = \sum_{i = 1}^{\numWaypoints - 1} \frac{|\subpathVessel_i|}{\sqrt{(\waypointOrigX{i+1} - \waypointOrigX{i})^2 + (\waypointOrigY{i+1} - \waypointOrigY{i})^2}}
\end{equation}
where $\pathVessel = \{\subpathVessel_1, \ldots, \subpathVessel_{\numWaypoints - 1}\} = \simulate(\vessel, \setWaypointsMod)$. The intuition is that the length of each sub-path $\subpathVessel_i$ (approximated by the number of its points) is normalized by the Euclidean distance between $\waypointOrigIndex{i} = (\waypointOrigX{i}, \waypointOrigY{i})$ and $\waypointOrigIndex{i+1} = (\waypointOrigX{i+1}, \waypointOrigY{i+1})$: the higher this value, the more unstable the sub-path is. The fitness is given by the sum of the unstability value of each sub-path.

\subsection{Search Operators}\label{sec:operators}

\approach uses classical simulated binary crossover and polynomial mutation operators with the default settings from PlatEMO~\cite{PlatEMO}-- the framework we used for the implementation of NSGA-II while adapting them for our specific problem's constraints.

In the crossover operation that recombines two individuals, we guarantee that the randomly selected crossover point occurs only between waypoints. This avoids splitting a waypoint, which could otherwise introduce too big changes to the waypoints.

The mutation operator performs changes at a coordinate level. It randomly changes the coordinate with a specified default probability from PlatEMO. While performing mutation, we check whether the constraint given by the solution encoding (see \Cref{sec:solution}) is satisfied, i.e., consecutive waypoints must be at least at a distance of \rSwitch; otherwise, the mutation is not applied.

\section{Empirical Evaluation Design}\label{sec:experimentDesign}

\subsection{Research Questions}\label{subsec:researchQuestions}
We aim to study the effectiveness of the \approach (with its three different seeding strategies) compared to a simple baseline, i.e., Random Search (\randomSearch). Moreover, we aim to study the characteristics of unexpected paths followed by AVs in response to waypoints generated by \approach. 
Consequently, we would like to answer the following research questions (RQs): 

\textbf{RQ1: How effective is \approach compared to random search (\randomSearch)?}
This RQ studies whether the problem we study is complex enough to require the sophisticated multi-objective algorithm NSGA-II or if it can be solved with a simple baseline technique, i.e., \randomSearch.
To this end, we compared \randomSearch with the three versions of \approach (i.e., for the three seeding strategies) with three different case studies.
To perform the analysis, we performed a pairwise comparison between \randomSearch and each of the \approach variations utilizing statistical tests.

\textbf{RQ2: How do different seeding strategies affect the effectiveness of the \approach?}
With this RQ, we aim to answer how the different seeding strategies affect \approach's effectiveness. To answer this RQ, we compared the effectiveness of the three seeding strategies, \approachSeed, \approachComb and \approachRnd, for each case study by utilizing statistical tests.

\textbf{RQ3: How good is \approach in generating paths leading to different types of unexpected paths?}
This RQ studies the different path types generated by the three versions of \approach. Studying this is important for understanding the characteristics of different unexpected paths, as such characterization helps determine which adaptations are required for each path type. To answer this RQ, we classified each case study and search approach's sub-paths to get insight into the types of paths generated and differences in the different case studies. 

\subsection{Setup and Execution} \label{subsec:setup}
\subsubsection{Case studies}
To evaluate \approach, we selected the Marine Systems Simulator (MSS) toolbox, which is designed in Matlab/Simulink to implement mathematical models of marine systems~\cite{MSSsimulator, perez2006overview}. MSS integrates various toolboxes to provide a comprehensive platform simulating different vessels. The simulator enables users to simulate scenarios such as dynamic positioning, autopilot design, and other marine operations. The simulator modular structure allows for flexibility in implementation, making it a practical tool for testing and rapid prototyping AVs, given that doing so on a real AV is expensive. MSS has been used in previous works to simulate ships, generating data for analysis~\cite{DARUS-2905_2022}, or testing marine operations~\cite{wang2021parameter}.

Our case studies are three AVs covering underwater and overwater AVs with different environments and a set of initial waypoints \setWaypointsOrig. 
Specifically, the selected AVs are the waypoint-guided AVs in MSS and comprise of one overwater tanker system, \mariner, and two underwater AVs, \resmus and \nspauv. 
The number of waypoints in \setWaypointsOrig (i.e, \numWaypoints) is 6 for \mariner, and 7 for \resmus and \nspauv, provided by the MSS initial configuration of the \setWaypointsGeneric.

\subsubsection{Search algorithms}
We implemented \approach using the evolutionary multi-objective optimization platform PlatEMO~\cite{PlatEMO}.
We utilized their implementation of NSGA-II~\cite{deb2002fast} and adapted their crossover operator to our operator described in \Cref{sec:operators}. Moreover, we replaced the default population generation of PlatEMO with the three different seeding strategies inside NSGA-II, resulting in three variations of \approach. As explained in \Cref{sec:initial_pop}, the specific seeding approaches are
\begin{inparaenum}[(i)]
\item a fully seeded initial population (\approachSeed), 
\item a combination of seeded and randomly generated waypoints (\approachComb), and 
\item a fully randomly generated initial population (\approachRnd).
\end{inparaenum}
To guide the search to \setWaypointsGeneric close to \setWaypointsOrig, all the initial populations of the different approaches contain one individual that is equal to \setWaypointsOrig.

For our baseline algorithm, Random Search (\randomSearch), we generate the same number of individuals as for \approach such that the total number of fitness evaluations is equal for \approach and \randomSearch to ensure a fair comparison. The initial population for \randomSearch is also generated randomly, as discussed in \Cref{sec:solution}.

\subsubsection{Parameter Settings}
By following the guide of Arcuri and Briand~\cite{ArcuriICSE11}, each variation of \approach and \randomSearch was run 30 times to account for the randomness of the search algorithms. Given that there are four approaches (the three seeding strategies of \approach and \randomSearch), in total we conducted 120 runs for each case study.

Due to the long execution time of the MSS simulator, an initial preliminary experiment was conducted to choose the most appropriate parameter settings, which involved analyzing the convergence of the Pareto Front. 
The population size was set to 10, and the maximum number of generations to 1000, resulting in a total number of 10,000 fitness evaluations. 
The default mutation operator and our slightly adapted crossover operator (see \Cref{sec:operators}) were used with default settings from PlatEMO. Specifically, the probability of crossover and the expectation of the number of mutated variables were both set to 1, whereas the distribution index of crossover and mutation were both set to 20.

The hyperparameter \deltaSearchSpace that defines the search space (see \Cref{sec:solution}) was set to 400 and 150 for the overwater AV and underwater AVs, respectively. These values were set based on the AVs' characteristics, such as AV length and original distance between waypoints.

\subsection{Evaluation Metrics} \label{subsec:evaluationmetrics}
\subsubsection{Quality indicator}
To assess the effectiveness of each approach, we employed the \textit{hypervolume} (HV) metric~\cite{ShangTEC2021} as our quality indicator. 
HV measures the volume of the objective space covered by the Pareto front with respect to a reference point, which is chosen based on the worst value observed across all approaches and experiments.
A higher HV value indicates better coverage and, thus, better performance and diversity of an approach.

\subsubsection{Statistical tests}\label{subsec:statisticalTests}
To compare the performance of the search approaches in RQ1 and RQ2, we employed the Mann-Whitney U-test for statistical significance and Vargha Delaney \Atwelve to measure the effect size following the guide~\cite{ArcuriICSE11}. Specifically, we compared each variation of \approach among themselves and with \randomSearch, resulting in a total of six comparisons for each AV.

The Mann-Whitney U-test gives a p-value that assesses wheth\-er there is a statistical significant difference between two approaches \textit{A} and \textit{B} in terms of HV.
If the p-value is above the significance level \signLevel ($\signLevel = 0.05$ in this work), it means there is \textit{No statistically significant Difference} (\noDifference). Instead, if the p-value is below \signLevel, there is a statistically significant difference. In this case, we check the effect size \Atwelve that tells the probability that one algorithm outperforms the other: if $\Atwelve > 0.5$, it means that \textit{A} is significantly better than \textit{B}; otherwise, if $\Atwelve < 0.5$, \textit{B} is significantly better than \textit{A}. We also check the strength of the effect size following the classification in~\cite{mangiafico2016}: \neglibleStrength when \Atwelve $\in$ (0.44, 0.56), \smallStrength when \Atwelve $\in$ (0.34,0.44] or \Atwelve $\in$ [0.56,0.64), \mediumStrength when \Atwelve $\in$ (0.29,0.34] or \Atwelve $\in$ [0.64,0.71), and \largeStrength when \Atwelve $\in$ [0,0.29] or \Atwelve $\in$ [0.71, 1].

\subsubsection{Classification of sub-paths and paths}\label{sec:classificationPaths}

\begin{figure}[!tb]
\centering
\begin{subfigure}[b]{0.49\columnwidth}
\centering
\includegraphics[width=\linewidth]{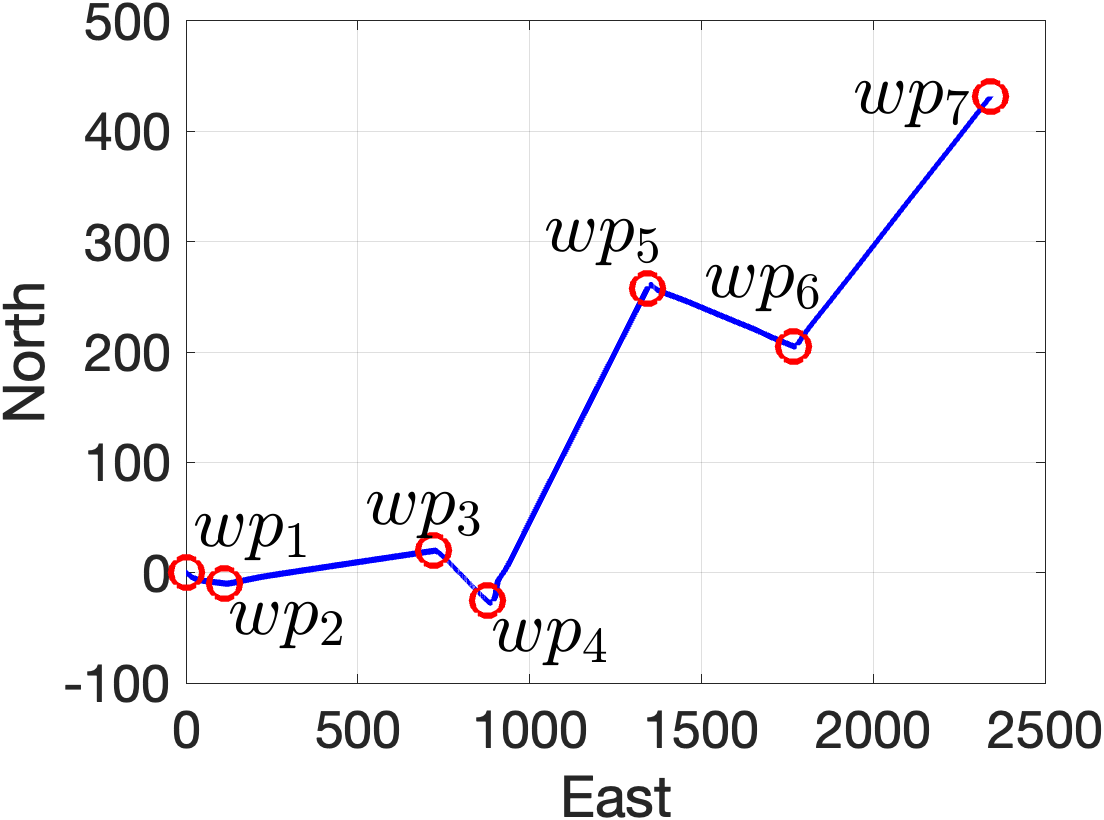}
\caption{Stable path}
\label{fig:exStraightExample2d}
\end{subfigure}
\hfill
\begin{subfigure}[b]{0.49\columnwidth}
\centering
\includegraphics[width=\linewidth]{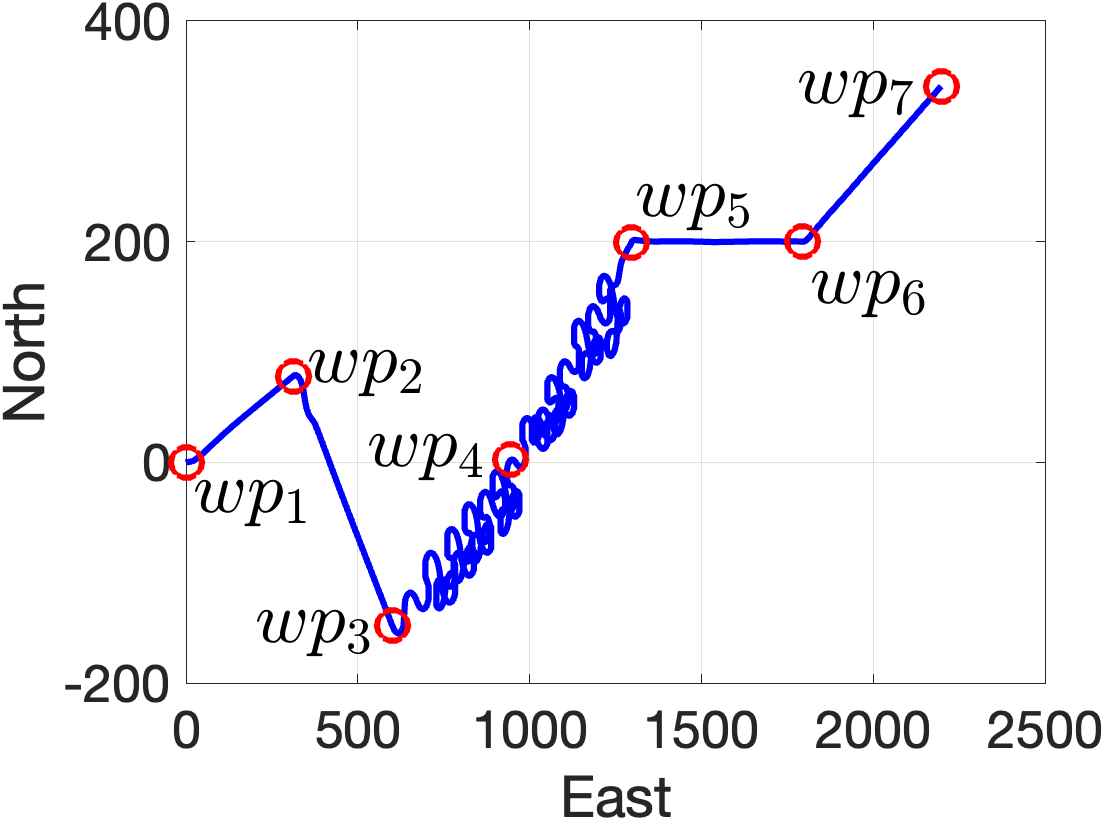}
\caption{Unstable path}
\label{fig:exCurlyPath2d}
\end{subfigure}
\begin{subfigure}[b]{0.49\columnwidth}
\centering
\includegraphics[width=\linewidth]{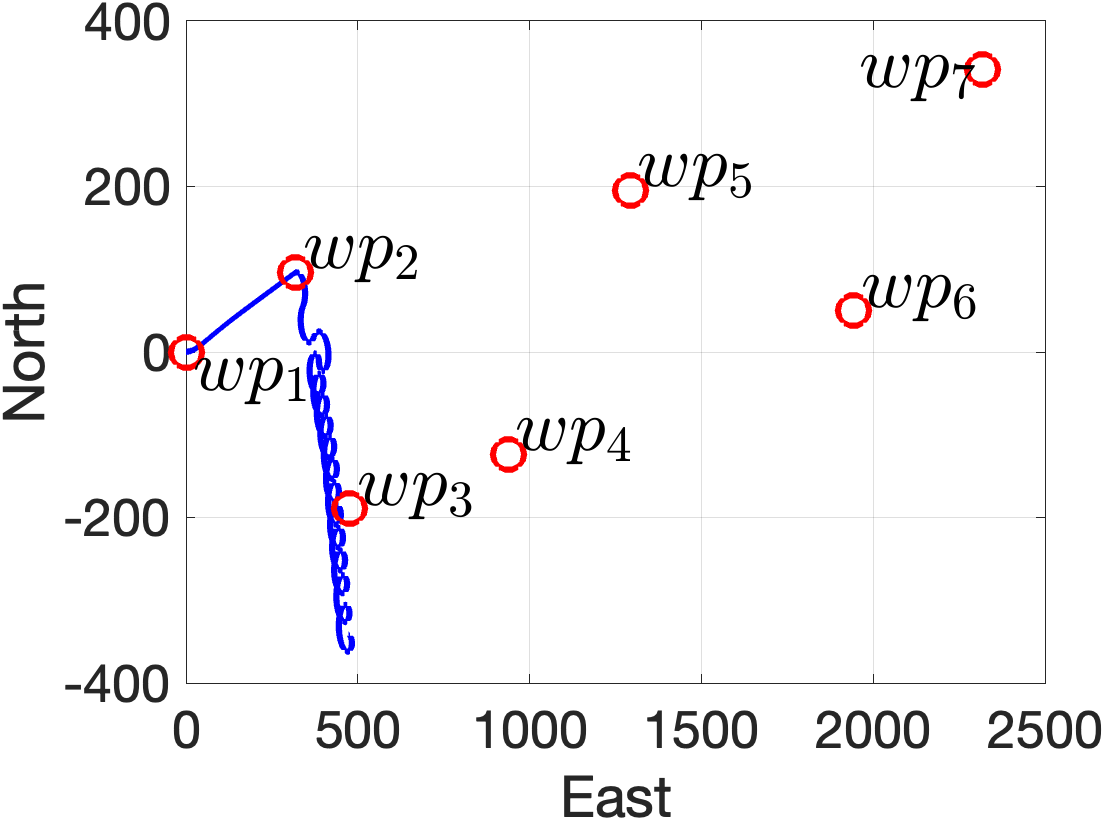}
\caption{Missing path}
\label{fig:exMissingPath2d}
\end{subfigure}
\caption{Example of paths of \resmus (North-East view) with different sub-path types: {\it stable}, {\it unstable}, and {\it missing}. A waypoint is shown as a red circle and labeled as \waypointGenericIndex{i} for $i_{\mathit{th}}$ waypoint. A path consists of waypoints from \waypointGenericIndex{1} to \waypointGenericIndex{7}. \Cref{fig:exCurlyPath2d} shows the AV's path is unstable between some waypoints. \Cref{fig:exMissingPath2d} shows the AV missing several waypoints.}
\label{fig:exampleTypesOfPaths2d}
\end{figure}

For RQ3, we are interested in studying the types of generated paths. We have defined three categories, \textit{stable}, \textit{unstable}, and \textit{missing}. The three sub-path types are shown in \Cref{fig:exampleTypesOfPaths2d}.
The stability/unstabilty of an AV can be observed by measuring how the angles of the AV change, such as rolling from side to side.
To detect this unstable behavior, we calculate the autocorrelation of the axis measurements to quantify the instability and differentiate between a stable measurement and an unstable one by calculating the number of peaks and the magnitude of each peak. We counted one angle as unstable if it had more than one threshold peak above 0.1~\cite{schober2018correlation}. With this, we can differentiate between a stable and an unstable sub-path based on how many axes are stable. Specifically:
\begin{inparaenum}
\item \textit{stable}: all of the angles are stable,
\item \textit{unstable}: at least one of the angles is unstable,
and
\item \textit{missing}: the vessel fails to reach a waypoint.
\end{inparaenum}

Using the previous classification sub-paths, we have a metric for describing the full path. Specifically, given the $\numWaypoints-1$ sub-paths of a path, a full path can have $2^{\numWaypoints-1} + 2 (\numWaypoints - 2)$ different categorizations.

\section{Results and Analyses}\label{sec:results}

\subsection{RQ1 -- Comparison with Random Search}\label{sec:rq1results}
\begin{figure}[!t]
\centering
\begin{subfigure}[b]{0.32\columnwidth}
\centering
\includegraphics[width=\textwidth]{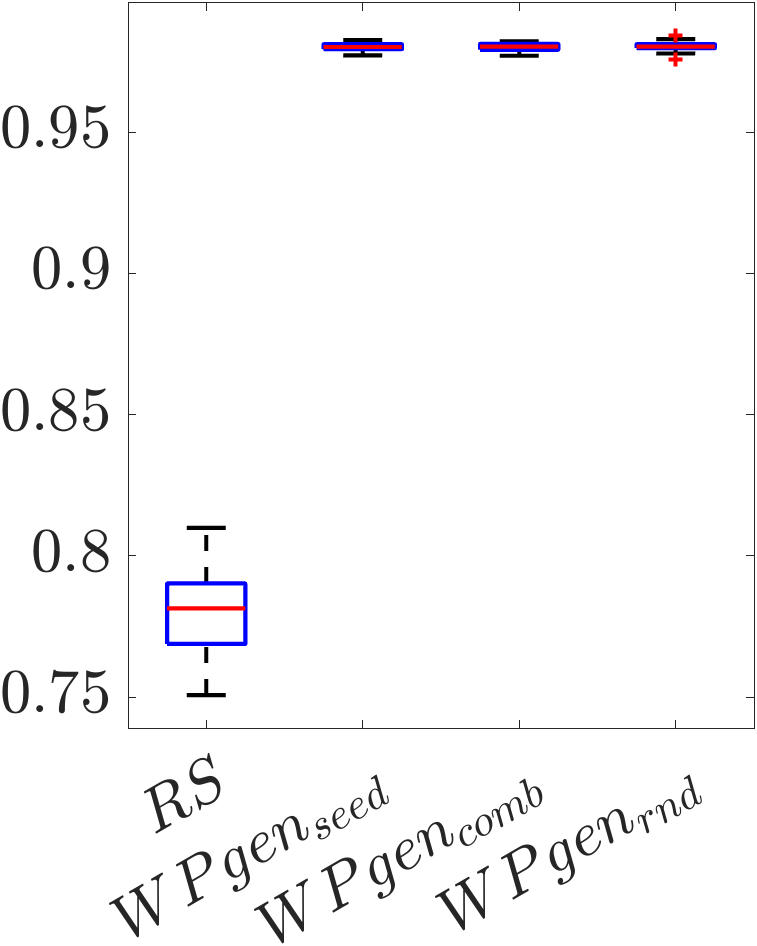}
\caption{\mariner}
\label{fig:HVboxPlotMariner}
\end{subfigure}
\begin{subfigure}[b]{0.32\columnwidth}
\centering
\includegraphics[width=\textwidth]{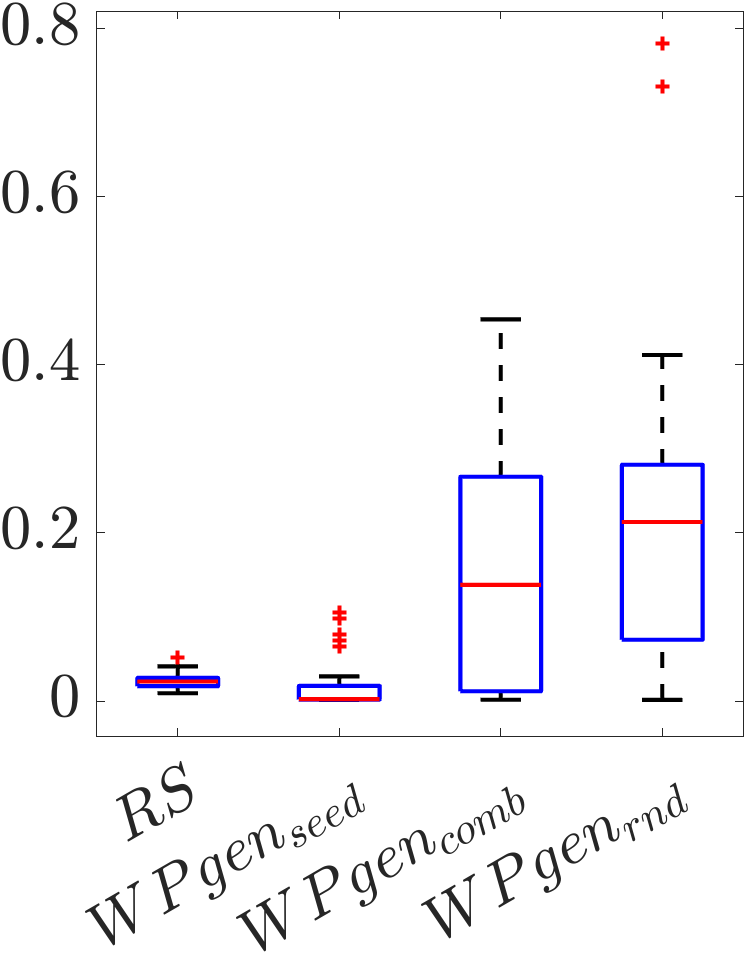}
\caption{\resmus}
\label{fig:HVboxPlotRemus100}
\end{subfigure}
\begin{subfigure}[b]{0.32\columnwidth}
\centering
\includegraphics[width=\textwidth]{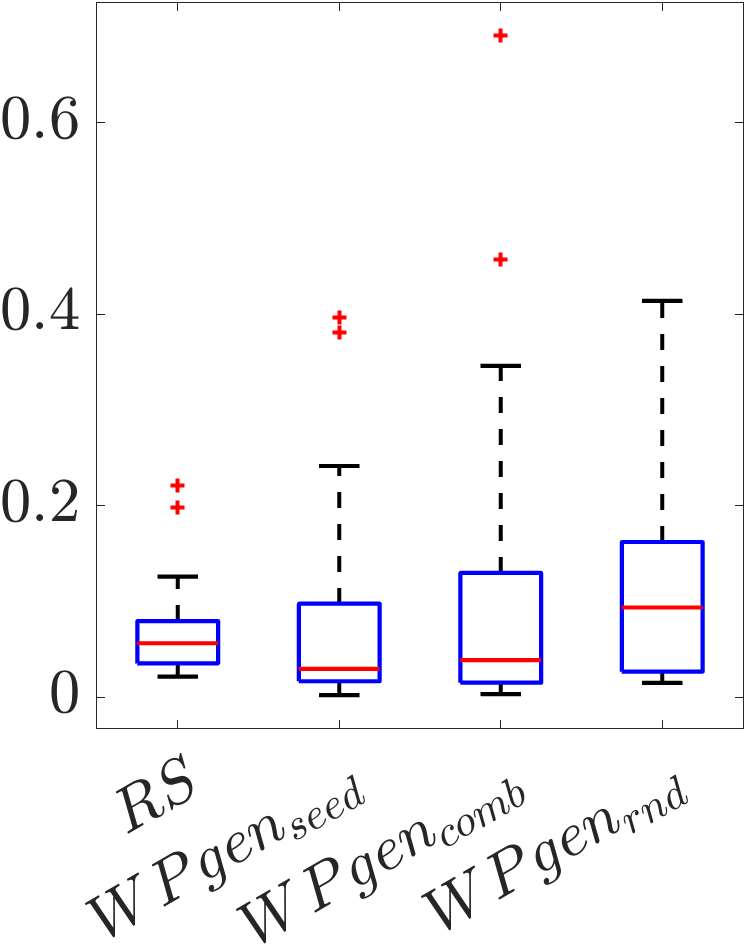}
\caption{\nspauv}
\label{fig:HVboxPlotNspauv}
\end{subfigure}
\caption{RQ1 and RQ2 -- Distribution of the calculated HV metric for each autonomous vessel and approach}
\label{fig:HVboxplotsAllships}
\end{figure}

\Cref{tab:RQ1RQ2statisticalTests} reports the results of the statistical tests when comparing the three variations of \approach with \randomSearch.
\begin{table}[!tb]
\centering
\caption{RQ1 and RQ2 -- Results of the statistical comparison between \approachSeed, \approachComb, \approachRnd, and \randomSearch, in terms of HV. \noDifference indicates no significant difference (i.e., p-value $\geq 0.05$). The \textit{Best Approach} column identifies the approach that is significantly better than the other (i.e., p-value $< 0.05$). The \textit{Strength} column indicates the strength of the significance based on the \Atwelve effect size, as described in \Cref{subsec:statisticalTests}}
\label{tab:RQ1RQ2statisticalTests}
\begin{tabular}{lllll}
\toprule
\multirow{2}{*}{AV \vessel} & \multicolumn{2}{c}{Compared approaches} & \multirow{2}{20pt}{Best approach} & \multirow{2}{*}{Strength} \\
\cmidrule{2-3} 
& A & B & & \\
\midrule
\multirow{6}{*}{\mariner} & \randomSearch & \approachSeed & \approachSeed & \largeStrength \\ 
& \randomSearch & \approachComb & \approachComb & \largeStrength \\ 
& \randomSearch & \approachRnd & \approachRnd & \largeStrength \\ 
& \approachSeed & \approachComb & \noDifference & \noDifference\\ 
& \approachSeed & \approachRnd & \noDifference & \noDifference\\
& \approachComb & \approachRnd & \noDifference & \noDifference\\ 
\midrule
\multirow{6}{*}{\resmus} & \randomSearch & \approachSeed & \randomSearch & \largeStrength\\ 
& \randomSearch & \approachComb & \approachComb & \largeStrength\\ 
& \randomSearch & \approachRnd & \approachRnd & \largeStrength \\ 
& \approachSeed & \approachComb & \approachComb & \largeStrength \\ 
& \approachSeed & \approachRnd & \approachRnd & \largeStrength \\ 
& \approachComb & \approachRnd & \noDifference & \noDifference\\ 
\midrule
\multirow{6}{4em}{\nspauv} & \randomSearch & \approachSeed & \noDifference & \noDifference\\ 
& \randomSearch & \approachComb & \noDifference & \noDifference\\ 
& \randomSearch & \approachRnd & \noDifference & \noDifference\\ 
& \approachSeed & \approachComb & \noDifference & \noDifference \\ 
& \approachSeed & \approachRnd & \noDifference & \noDifference \\ 
& \approachComb & \approachRnd & \noDifference & \noDifference \\ 
\bottomrule
\end{tabular}
\end{table}
For each AV, the comparison with \randomSearch is shown in rows where the \textit{Approach A} column has a value of \randomSearch. The fourth column indicates which of the two approaches performs significantly better, followed by the strength category computed based on \Atwelve (see \Cref{subsec:statisticalTests}).
If no conclusion can be drawn from the statistical test (i.e., p-value $<0.05$), \noDifference is written to indicate no significant difference between the two compared approaches. 
Moreover, the descriptive statistics of the results are shown as the boxplots showing the distributions of the HV values for each AV and approach in \Cref{fig:HVboxplotsAllships}.

For \mariner, when analyzing the distributions of the HV values in \Cref{fig:HVboxPlotMariner}, we can see that \randomSearch is worse than the three versions of \approach. Moreover, \randomSearch has more distribution variance than \approach, indicating that the effectiveness of \randomSearch is unreliable. Furthermore, the results of statistical tests in \Cref{tab:RQ1RQ2statisticalTests} show that each variation of \approach is significantly better than \randomSearch with \largeStrength strength.

For \resmus, \Cref{fig:HVboxPlotRemus100} shows that \approachComb and \approachRnd perform better than \randomSearch; however, the variance in their distribution is high. \approachSeed, on the other hand, seems to perform worse than \randomSearch. Both \randomSearch and \approachSeed have less variances in their distributions. These observations are confirmed by the results of the statistical tests in \Cref{tab:RQ1RQ2statisticalTests}, where we can see that \approachComb and \approachRnd perform significantly better than \randomSearch with \largeStrength strength. However, \randomSearch performs significantly better than \approachSeed with \largeStrength strength. 

For \nspauv, when comparing the distributions of the HV values in \Cref{fig:HVboxPlotNspauv} between \randomSearch and each of three variations of \approach, we do not see clearly which approach performs the best. This is further confirmed by the results of statistical tests reported in \Cref{tab:RQ1RQ2statisticalTests} showing no significant differences among any pair of approaches.

\conclusion{\textbf{Conclusion for RQ1}: \approach either performs significantly better than \randomSearch or shows no significant differences depending on the autonomous vessel, except \resmus, where \randomSearch is significantly better than \approachSeed. Thus, we can conclude that our problem is complex enough to require a guided algorithm such as NSGA-II, except \nspauv.}

\subsection{RQ2 -- Comparison of Seeding Strategies}\label{sec:rq2results}
\Cref{fig:HVboxplotsAllships} reports the distribution of the HV values for the three version of \approach (i.e., \approachSeed, \approachComb, and \approachRnd), while \Cref{tab:RQ1RQ2statisticalTests} reports the results of their statistical comparison.

For \mariner, when comparing the distributions of the three variations of \approach in \Cref{fig:HVboxPlotMariner}, we do not see any differences; this is further confirmed by the results of statistical tests in \Cref{tab:RQ1RQ2statisticalTests}, which show no significant differences among the three approaches.

For \resmus, when looking at the distributions of the three variations of \approach in \Cref{fig:HVboxPlotRemus100}, we see no clear difference between \approachComb and \approachRnd; this is further confirmed by the results of statistical tests in \Cref{tab:RQ1RQ2statisticalTests}, which show no differences between the two approaches. Instead, in \Cref{fig:HVboxPlotRemus100}, \approachSeed seems worse than the other two approaches; indeed, results of the statistical tests in \Cref{tab:RQ1RQ2statisticalTests} show that \approachSeed is significantly worse than \approachComb and \approachRnd with \largeStrength strength. Recall from RQ1 (see \Cref{sec:rq1results}) that \approachSeed was also outperformed by \randomSearch for \resmus.

For \nspauv, when looking at the distributions of three approaches in \Cref{fig:HVboxPlotNspauv}, we do not see clear differences; this is confirmed by the results in \Cref{tab:RQ1RQ2statisticalTests} that report no significant differences. 

\conclusion{\textbf{Conclusion for RQ2}: 
The suitability of a seeding strategy depends on the AV \vessel being tested. For instance, for \mariner and \nspauv, it does not matter which strategy to use, whereas, for \resmus, either \approachComb or \approachRnd is the best strategy.}

\subsection{RQ3 -- Paths Classification}\label{sec:rq3results}

\Cref{tab:AvgCountOfDifferentPaths} reports, for each approach and AV, the percentage of different sub-paths calculated from all solutions on the Pareto front (average across 30 runs).
\begin{table}[!tb]
\centering
\caption{RQ3 -- Percentage (across 30 runs) of different sub-paths (i.e., stable, unstable, and missing) from all the solutions on a Pareto Front }
\begin{tabular}{llrrr}
\toprule
\multirow{2}{*}{Approach} & \multirow{2}{4em}{Type of path} & \multicolumn{3}{c}{AV \vessel} \\
\cmidrule{3-5} 
& & \mariner & \resmus & \nspauv \\
\midrule
\multirow{3}{*}{\approachSeed} & Stable & 16.89\% & 74.72\% & 79.84\% \\ 
& Unstable & 83.11\% & 20.17\% & 18.20\% \\ 
& Missing &0 & 5.12\% & 1.95\% \\ 
\midrule
\multirow{3}{*}{\approachComb} & Stable & 17.57\% & 68.98\% & 68.88\% \\ 
& Unstable & 82.43\% & 24.47\% & 25.40\% \\ 
& Missing &0 & 6.54\% & 5.72\% \\ 
\midrule
\multirow{3}{*}{\approachRnd} & Stable & 18.72\% & 68.12\% & 69.80\% \\ 
& Unstable & 81.28\% & 25.23\% & 25.30\% \\ 
& Missing & 0 & 6.65\% & 4.90\% \\ 
\bottomrule
\end{tabular}
\label{tab:AvgCountOfDifferentPaths}
\end{table}

For \mariner, the percentages of stable sub-paths across the three approaches are similar: 16.89\%, 17.57\%, and 18.72\% for \approachSeed, \approachComb, and \approachRnd, respectively. Similarly, the percentages for unstable paths are also very close: 83.11\%, 82.43\%, and 81.28\%. None of the three approaches found any missing sub-paths. These results suggest that all three approaches perform similarly in identifying unstable sub-paths for \mariner. Furthermore, the higher percentage of unstable sub-paths indicates that it was relatively easier for any approach to detect unstable paths for \mariner.

For \resmus, \approachSeed identified a slightly higher percentage of stable sub-paths (74.72\%) compared to \approachComb (68.98\%) and \approachRnd (68.12\%). In contrast, for unstable sub-paths, \approachSeed had a slightly lower percentage (20.17\%) compared to \approachComb (24.47\%) and \approachRnd (25.23\%). Regarding missing sub-paths, all three approaches exhibited similar percentages: 5.1\% for \approachSeed, 6.54\% for \approachComb, and 6.65\% for \approachRnd. These results suggest that \approachComb and \approachRnd are slightly better than \approachSeed at identifying unstable and missing sub-paths.

For \nspauv, the results are similar to those of \resmus. \linebreak
\noindent \approachSeed achieved the highest percentage of stable sub-paths (79.84\%) compared to \approachComb (68.88\%) and \approachRnd (69.80\%). \approachComb (25.4\%) and \approachRnd (25.3\%) had higher percentages of unstable sub-paths than \approachSeed (18.2\%). Similarly, \approachComb (5.72\%) and \approachRnd (4.92\%) recorded higher percentages of missing sub-paths than \approachSeed (1.95\%). Once again, these results suggest the better suitability of \approachComb and \approachRnd for identifying unstable and missing sub-paths for \nspauv.

As explained in \Cref{sec:classificationPaths}, each path followed by a vessel can be classified by considering the combination of the classifications of its sub-paths; at most, there are $2^{\numWaypoints-1} + 2 (\numWaypoints - 2)$ possible types of paths. \Cref{tab:RQ3AvgCountOfUniquePaths} indicates the average percentage of unique full paths generated per approach for each AV across the 30 runs.
\begin{table}[!tb]
\centering
\caption{RQ3 -- Average (across 30 runs) percentage of unique full paths from all the solutions on a Pareto Front. The percentage is based on the maximum number of possible paths}
\label{tab:RQ3AvgCountOfUniquePaths}
\begin{tabular}{lrrr}
\toprule
\multirow{2}{4em}{Approach} & \multicolumn{3}{c}{AV \vessel} \\
\cmidrule{2-4} 
& \mariner  & \resmus  & \nspauv\\
\midrule 
\approachSeed & 9.65\% & 23.24\% & 36.35\% \\ 
\approachComb & 12.00\% & 30.45\% & 70.62\% \\ 
\approachRnd & 13.15\% & 37.88\% & 74.81\%\\
\bottomrule
\end{tabular}
\end{table}
For \mariner, \approachRnd manages to generate, on average, slightly more unique paths (i.e.,  13.15\%) than the other approaches, i.e., 12.00\% for \approachComb and 9.65\% for \approachSeed. A similar trend is observed for \resmus and \nspauv, where the average percentage of unique paths is highest for \approachRnd and lowest for \approachSeed. Such results are expected since total random seeding will result in more unique paths, whereas fully seeding the initial population causes the approach to focus on those waypoints. At the same time, the combination of both approaches produces results that fall in the middle. Another observation is that \mariner, generally across all approaches, has the lowest average percentage of unique paths, followed by \resmus, whereas \nspauv has the highest average number of unique paths. This result suggests that the number of unique paths will vary depending on the AV.

\conclusion{\textbf{Conclusion for RQ3}:  For underwater autonomous vessels (\resmus and \nspauv), \approachComb and \approachRnd, which incorporate random individuals in the initial population, generate more paths with unstable and missing sub-paths. For overwater autonomous vessels (\mariner), all approaches perform similarly in identifying unstable and missing sub-paths. Overall, for any autonomous vessel type, approaches with random initial populations are recommended. Moreover, \approachSeed identifies fewer unique paths than \approachComb and \approachRnd, with the number of unique paths varying across vessels.}

\section{Discussion} \label{sec:discussion}

\subsection{Research and Practical Implications} \label{subsec:researcptactucal}

Based on the results from RQ1, we found that all three variations of \approach either perform significantly better than \randomSearch or show no significant differences, depending on the AV. Similarly, results for RQ2 showed differences among the approaches for different AVs. These results suggest that the selection of a particular \approach depends on the AV being tested.
For \mariner, while there were no significant differences among the three variations of \approach, all were significantly better than \randomSearch, indicating that any variation of \approach is sufficient for testing \mariner. For \resmus, \approach proved to be the best, making it the natural choice for testing. 
For \nspauv, there were no significant differences between \randomSearch and the three variations of \approach, meaning that even \randomSearch is sufficient for testing \nspauv. Although we now have preliminary indications for selecting approaches, it raises further questions about why a specific approach is better for a particular AV. This requires additional empirical evaluations that consider more AV characteristics. Such a study is part of our future work.

From RQ3, we found that approaches incorporating randomness in the initial population outperformed those that fully seed the initial population for both underwater and overwater AVs. Therefore, such approaches—\approachComb and \approachRnd—are recommended for identifying waypoints that should trigger adaptations in self-adaptive AVs. From RQ3, we also found that \approachSeed generally identified fewer unique full paths than \approachComb and \approachRnd. These results were expected, as \approachSeed is designed to enable the search algorithm to focus around the seeded waypoints. Depending on the context, a tester may want to focus only on a given set of waypoints; in that case, \approachSeed is recommended. However, it naturally lacks the diverse coverage the other two approaches provide. On the other hand, if a tester seeks more diverse paths, \approachComb or \approachRnd are recommended.

In general, we conclude that \approach can find unstable paths close to the original trajectory \setWaypointsOrig. Thus, our results bring new knowledge about the importance of \setWaypointsGeneric generation and analysis of AVs. Furthermore, the waypoints that result in unstable paths can be utilized for different purposes. First, these waypoints can be used to trigger different adaptations in self-adaptive AVs to verify whether correct adaptations are triggered and implemented. Second, further analysis can be conducted to identify any patterns that explain why these \setWaypointsGeneric specifically result in unstable paths and how such \setWaypointsGeneric can be avoided in the future. Finally, our methods for categorizing the sub-paths can be further utilized to create more advanced fitness functions to further enhance the performance of \approach. 

Our approach \approach can be used to pre-evaluate a set of waypoints \setWaypointsGeneric (constructed as minor variations of the original ones \setWaypointsOrig), and determine the safety of the AV. Detection of a probable unstable path can be utilized as feedback to the system, letting the AV adjust its controllers to get a smoother and safer path. A set of waypoints \setWaypointsGeneric that lead to a path having unstable or missing sub-paths, can be utilized for different AV navigation software to compare and determine if it is better at handling \setWaypointsGeneric. Navigation software can also utilize the characteristics of the \setWaypointsGeneric to recognize critical situations that require adaptation.

\subsection{Threats to Validity} \label{subsec:threats}


\noindent{\bf Internal validity.}
An internal validity threat is related to the parameters of the selected algorithm, i.e., NSGA-II, which is the core part of the three variations of \approach. To mitigate this threat, we used the default crossover and mutation operators with default parameter settings from PlatEMO. These parameters were then kept the same for the three variations of \approach for a fair comparison. One could argue that we selected only one algorithm, i.e., NSGA-II; however, it is widely used in search-based software engineering, and we naturally plan to investigate additional algorithms in the future. \looseness=-1

\smallskip
\noindent{\bf External validity}
We experimented with three AVs of different characteristics in a fixed environment. Thus, the results may not generalize to other AVs or environments. This highlights the need for additional empirical evaluations, which, along with exploring other search algorithms, will form the basis of our future studies.

\smallskip
\noindent{\bf Conclusion validity}
All search algorithms have inherent randomness. To address this, we executed each approach and \randomSearch 30 times, following a well-established guide~\cite{ArcuriICSE11}. Additionally, to ensure greater confidence in the results, we analyzed them using well-established statistical tests as recommended in the same guide.

\smallskip
\noindent{\bf Construct validity}
A concern is related to measures used to draw conclusions. For example, to automatically classify paths, we used autocorrelation; however, other metrics, such as zero-crossing or frequency analysis, also exist. We did not select these metrics because they were less precise and misclassified some stable paths as unstable during our exploratory study. Another concern relates to the stopping criteria, which in our context was defined as the number of fitness evaluations. Based on an exploratory study, this was set to 10,000 and kept consistent across all approaches.

\section{Related Work}\label{sec:related}

In this work, we proposed an approach to generate tests exposing cases in which an AV follows unstable paths. Such cases will be considered during the development of a self-adaptive AV in the context of our project (see \Cref{sec:realWorldContext}). To the best of our knowledge, this is the first approach proposed for this goal. On the other hand, in the self-adaptive community~\cite{selfAdaptiveFirstRoadMap,selfAdaptiveSecondRoadMap}, some works have been proposed to test the self-adaptive system itself. H{\"a}nsel et al.~\cite{Hansel2015} proposed a model-based testing approach for MAPE-K that relies on architectural runtime models (RTM). Specifically, the approach checks at runtime the conformance between the running self-adaptive system and the oracle RTM. Eberhardinger et al.~\cite{Eberhardinger2014TTS} proposed an approach based on model checking to generate test cases for self-adaptive systems. In order to make the approach scalable, they simplify the system using the Corridor Enforcing Infrastructure (CEI) architectural pattern. Arcaini et al.~\cite{mbtSelfAdaptAMOST20} also proposed an approach based on model checking to derive test cases from self-adaptive ASMs, a formal notation for the analysis of self-adaptive systems. Fredericks et al.~\cite{Fredericks13,Fredericks14}, instead, proposed online testing techniques able to better adjust to the uncertainty of the environment.

Our work is also related to Search-Based Software Engineering (SBSE), which formulates software engineering problems as optimization problems and solves them using search algorithms, such as single-objective and multi-objective search algorithms~\cite{SSBSE}. Since its inception, SBSE has been applied to various software engineering sub-areas, e.g., requirements engineering, modeling, testing, and debugging~\cite{SBSETrends}. Within SBSE, Search-Based Testing (SBT) focuses specifically on testing software systems using search algorithms, which is the focus of this paper. SBT has been applied to testing a variety of systems, including autonomous driving systems~\cite{ADSSBT}, autonomous robots~\cite{SBTFramework}, unmanned aerial vehicles (UAVs)~\cite{UAVSBT}, automated warehouses~\cite{SBTCPS}, and sports-related cyber-physical systems~\cite{SBTCPS}. Despite the applications of SBT in various domains, its application for testing the navigation software of AVs remains understudied. To this end, this paper presents an application of SBT to generate tests (i.e., minor variations in the predefined waypoints) that should trigger adaptations in AVs.

Scenario-based testing with simulators is one of the key methods for testing AVs at design time~\cite{VVANavigation,ScenariNNTest,DLforNavigation,SMTwGaussian,RTforAVTest,SearchApproachMuliShipCollision}, which is also related to \approach. Most scenario-based testing works have primarily focused on collision avoidance. Furthermore, existing works often generate a limited number of scenarios with fixed settings, and such generation is often manual, as concluded in a literature review 
by ~\citeauthor{VVANavigation} ~\cite{VVANavigation}.
Nonetheless, the need for scenario-based testing is being been recognized, and recent works have explored more systematic and automated approaches to simulation-based testing for collision avoidance, such as using machine learning techniques~\cite{ScenariNNTest,DLforNavigation}, random testing~\cite{RTforAVTest}, or a search-based approach~\cite{SearchApproachMuliShipCollision}. Our work distinguishes itself from existing works by generating tests that should trigger adaptation in AVs. Specifically, we generate variations of predefined waypoints with SBT, so reducing the manual effort required to identify variations that could potentially impact the AV's ability to follow its intended path.

\section{Conclusions and Future Work}\label{sec:conclusion}
We presented a multi-objective search approach, \approach, to generate tests that should trigger adaptations in maritime autonomous vessels (AVs), where each test is a slight variation of the pre-defined waypoints that the AV must follow during navigation. The approach was further varied into three versions, each employing a different seeding strategy to study how initial population seeding affects performance. We compared the three variations with \randomSearch and with each other. Based on the results, we found that, in general, \approach either performs significantly better than \randomSearch or similarly. Moreover, the best-performing variation of the approach depends on the AV. In the future, we will extend \approach from several perspectives, including defining domain-specific mutation and crossover operators. Moreover, the empirical evaluation will be extended with more algorithms, different environmental setups, and similar systems, such as unmanned aerial vehicles. Finally, we also want to explore machine learning algorithms to detect unstable paths, gain insights into the different characteristics of an unstable path, and determine the root cause of the unstable paths. 

\textbf{Replication package}: For replicability, we make all our code, experimental scripts, results, and detailed plots available at Zenodo \url{https://zenodo.org/records/15093003}.

\section*{Acknowledgments}
K. Nylænder and S. Ali are supported by the Co-tester project (Project\# 314544), funded by the Research Council of Norway. S. Ali is also supported by the RoboSAPIENS project, funded by the European Commission’s Horizon Europe programme (Grant Agreement \# 101133807). P. Arcaini is supported by Engineerable AI Techniques for Practical Applications of High-Quality Machine Learning-based Systems Project (Grant Number JPMJMI20B8), JST-Mirai. Aitor Arrieta is part of the Systems and Software Engineering group of Mondragon Unibertsitatea (IT1519-22), supported by the Department of Education, Universities and Research of the Basque Country.  

\bibliographystyle{unsrtnat}
\bibliography{GECCO2025_waypoints_cameraReady} 

\end{document}